\documentstyle[11pt]{article}
\begin{document}
\title{Deflection of light by the screw dislocation in space-time}
\author{\it Miroslav Pardy\\
\it Department of Theoretical Physics and Astrophysics \\
\it Masaryk University \\
\it Kotl\'{a}\v{r}sk\'{a} 2, 611 37 Brno, Czech Republic\\
e-mail:pamir@physics.muni.cz}
\date{\today}
\maketitle
\vspace{50mm}

\begin{abstract}
We derive the light deflection caused by the screw dislocation in
space-time.
The derivation is based on the idea that space-time is a medium
which can be deformed by gravity and that the deformation of space-time
is equivalent to the existence of gravity.
\end{abstract}

\newpage

\section{Introduction}

\hspace{3ex}
\baselineskip 15pt

There is a possibility that during the big bang, supernova explosion,
gravitational collaps, collisions of the high-energy elementary particles
and so on, the dislocations in space-time are
created. In this article we derive the deflection of light caused by
the screw dislocation in space-time.

In order to derive such deflection of light,
it is necessary to explain the origin of metric in the Einstein
theory of gravity.

Einstein gives no explanation of the origin of the
metrics, or, metrical tensor. He only introduces the Riemann geometry
as the basis for the general relativity [1].
He "derived" the nonlinear
equations for the metrical tensor [2] and never
explained what is microscopical origin of the metric of space-time.
Einstein supposed that it is adequate that
the metric follows from differential equations as their solution.
However, the metric has an microscopical origin similarly to
situation where the phenomenological thermodynamics
has also the microscopical and statistical origin.

The first question we ask, is,  what is the
microscopical origin of the metric of space-time.
We postulate that the origin of metric is the specific
deformation of space-time continuum. We take the idea from
the mechanics of continuum and we apply it to
the space-time medium. The similar approach can be found
in the Tartaglia article and author eprint, [3], where
space-time is considered as a deformable medium.

The mathematical description of the three dimensional deformation is
given for instance in [4]. The fundamental quantity is the
tensor of deformation expressed
by the relative displacements $u^{i}$ as follows:

\begin{equation}
u_{ik} = \left(\frac {\partial u_{i}}{\partial x^{k}} +
\frac {\partial u_{k}}{\partial x^{i}} + \frac {\partial u^{l}}{\partial
x^{i}} \frac {\partial u_{l}}{\partial x^{k}}\right);\quad i, k = 1, 2, 3.
\label{1}
\end{equation}

The last definition can be generalized to the four dimensional situation
by the following relation:

\begin{equation}
u_{\mu\nu} = \left(\frac {\partial u_{\mu}}{\partial x^{\nu}} +
\frac {\partial u_{\nu}}{\partial x^{\mu}} +
\frac {\partial u_{\alpha}}{\partial x^{\mu}}
\frac {\partial u^{\alpha}}{\partial x^{\nu}}\right); \quad
\mu, \nu = 0,1, 2,3,
\label{2}
\end{equation}
with $x^{0} = ct, x^{1} = x, x^{2} = y, x^{3} = z.$

 In order to establish the connection between metric $g_{\mu\nu}$ and
deformation expressed by the tensor of deformation, we write
for the metrical tensor $g_{\mu\nu}$ in squared space-time element

\begin{equation}
ds^{2} = g_{\mu\nu}dx^{\mu}dx^{\nu},
\label{3}
\end{equation}
the following relation

\begin{equation}
g_{\mu\nu} = (\eta_{\mu\nu} + u_{\mu\nu}),
\label{4}
\end{equation}
where

\begin{equation}
\eta_{\mu\nu} =
\left(\begin{array}{cccc}
1 & 0 & 0 & 0\\
0 & -1 & 0 & 0\\
0 & 0 & -1 & 0\\
0 & 0 & 0 & -1\\
\end{array}\right).
\label{5}
\end{equation}

Instead of work with the metrical tensor $g_{\mu\nu}$, we can work with
the tensor of deformation $u_{\mu\nu}$ and we can consider the general theory
of relativity as the four-dimensional theory of some
real deformable medium as a partner of the metrical theory.
First, let us test the deformation approach to the space-time in case of the
nonrelativistic limit.

\section{The nonrelativistic test}

The Lagrange function of a point particle with mass $m$ moving in
a potential $\varphi$ is given by the following formula [5]:

\begin{equation}
L = -mc^{2} + \frac {mv^{2}}{2} - m\varphi .
\label{6}
\end{equation}

Then, for a corresponding action we have

\begin{equation}
S = \int L dt = -mc \int dt \left(c - \frac {v^{2}}{2c} +
\frac {\varphi}{c} \right) ,
\label{7}
\end{equation}
which ought to be compared with $S = -mc\int ds$. Then,

\begin{equation}
ds = \left(c - \frac {v^{2}}{2c} +  \frac {\varphi}{c} \right) dt.
\label{8}
\end{equation}

With $ d{\bf x} = {\bf v}dt$ and neglecting higher derivative terms, we have

\begin{equation}
ds^{2} = (c^{2} + 2\varphi) dt^{2} - d{\bf x}^{2} =
\left(1 + \frac {2\varphi}{c^{2}}\right)c^{2}dt^{2} - d{\bf x}^{2}.
\label{9}
\end{equation}

The metric determined by this $ds^{2}$ can be be obviously related
to the $u_{\alpha}$ as follows:

\begin{equation}
g_{00} =  1 + 2\partial_{0}u_{0} +
\partial_{0}u^{\alpha}\partial_{0}u_{\alpha} =
1 + \frac {2\varphi}{c^{2}}.
\label{10}
\end{equation}

We can suppose that the time shift caused by the potential is
small and therefore we can neglect the nonlinear term in the last
equation. Then we have

\begin{equation}
g_{00} =  1 + 2\partial_{0}u_{0}  = 1 + \frac {2\varphi}{c^{2}}.
\label{11}
\end{equation}

The elementary consequence of the last equation is

\begin{equation}
\partial_{0}u_{0} =  \frac {\partial u_{0}}{\partial (ct)} =
\frac {\varphi}{c^{2}},
\label{12}
\end{equation}
or,

\begin{equation}
u_{0} = \frac {\varphi}{c}t + const.
\label{13}
\end{equation}

Using $u_{0} = g_{00}u^{0}$, or, $u^{0} = g^{-1}_{00}u_{0} =
\frac {\varphi}{c}t$, we get with $const. = 0$ and

\begin{equation}
u^{0} = ct' - ct,
\label{14}
\end{equation}
the following result

\begin{equation}
t'(\varphi) = t(0)\left(1 + \frac {\varphi}{c^{2}}\right),
\label{15}
\end{equation}
which is the Einstein formula relating time in the zero
gravitational field to time in the gravitational potential
$\varphi$. The time interval $t(0)$ measured remotely is so called the
coordinate time and $t(\varphi)$ is local proper time.
The remote observe measures time intervals to be delated and
light to be red shifted.
The shift of light frequency corresponding to the gravitational
potential is, as follows [5].

\begin{equation}
\omega = \omega_{0}\left(1 + \frac {\varphi}{c^{2}}\right).
\label{16}
\end{equation}

The precise measurement of the gravitational spectral shift was made by Pound
and Rebka in 1960. They predicted spectral shift $\Delta\nu/\nu =
2.46 \times 10^{-15}$ [1].
The situation with red shift is in fact closed problem and no additional
measurement is necessary.

While we have seen that the red shift follows
from our approach immediately,
without application of the Einstein equations,
it is evident that the metrics
determined by the Einstein equations can be expressed by the tensor of
deformation. And vice versa, to the every tensor of
deformation the metrical tensor corresponds.

\section{The deflection of light by the screw dislocation}

According to [4] the screw deformation is defined by the tensor of deformation
which is in the cylindrical coordinates as
\begin{equation}
u_{z\varphi} = \frac {b}{4\pi r},
\label{17}
\end{equation}
where $b$ is the $z$-component of the
Burgers vector. The Burgers vector of the screw islocation
has components $b_{x} = b_{y} = 0, b_{z} = b$.

The postulation of the space-time as a medium enables to transfer the notions
of the theory of elasticity into the relativistic theory of space-time and
gravity. The considered transfer is of course the heuristical
operation, nevertheless the consequences are interesting.
To our knowledge, the problem, which we solve is new.

We know that the metric of the empty space-time is defined by the
coefficients in the relation:

\begin{equation}
ds^{2} = c^{2}dt^{2} - dr^{2} - r^{2} d\varphi^{2}  - dz^{2}.
\label{18}
\end{equation}

If the screw deformation is present in space-time, then the generalized metric
is of the form:
\begin{equation}
ds^{2} = c^{2}dt^{2} - dr^{2} - r^{2} d\varphi^{2} - 2u_{z\varphi}dzd\varphi
- dz^{2},
\label{19}
\end{equation}
or,

\begin{equation}
ds^{2} = c^{2}dt^{2} - dr^{2} - r^{2} d\varphi^{2} -
\frac {2b}{4\pi r}dzd\varphi - dz^{2}.
\label{20}
\end{equation}

The motion of light in the Riemann space-time is described by the equation
$ds = 0$. It means, that from the last equation the following
differential equation for photon follows:

\begin{equation}
0 = c^{2} -{\dot r}^{2} - r^{2} {\dot\varphi}^{2} -
\frac {b}{2\pi r}\dot z \dot\varphi - {\dot z}^{2}.
\label{21}
\end{equation}

Every parametric equations which obeys the last equation are equation of
motion of photon in the space-time with the screw dislocation.
Let us suppose that the motion of light is in the direction of the z-axis.
Or, we write

\begin{equation}
r = a;\quad {\dot z} = v.
\label{22}
\end{equation}
Then, we get equation of $\varphi$:

\begin{equation}
2 \pi a^{3} \dot\varphi^{2} + b v\dot\varphi = 2\pi a(c^{2} - v^{2}).
\label{23}
\end{equation}
We suppose that the solution of the last equation is of the form

\begin{equation}
\varphi = At.
\label{24}
\end{equation}
Then, we get for the constant $A$  the quadratic equation

\begin{equation}
2\pi a^{3}A^{2} + bvA + 2\pi a(v^{2} - c^{2}) = 0
\label{25}
\end{equation}
with the solution

\begin{equation}
A_{1/2} = \frac {-bv \pm \sqrt{b^{2}v^{2} - 16\pi^{2} a^{4}(v^{2} - c^{2})}}{4\pi a^{3}}.
\label{26}
\end{equation}

Using approximation $v \approx c$, we get that first root is approximately
zero and for the second root we get:

\begin{equation}
A \approx \frac {-bc}{2\pi a^{3}},
\label{27}
\end{equation}
which gives the function $\varphi$ in the form:

\begin{equation}
\varphi \approx \frac {-bc}{2\pi a^{3}}t.
\label{28}
\end{equation}

Then,  if $z_{2} - z_{1} = l$ is a distance between two points on the straight
line parallel with the axis of screw dislocation then,
$\Delta t = l/c$, $c$ being the velocity of light.
For the deflection angle $\Delta\varphi$, we get:

\begin{equation}
\Delta\varphi \approx \frac {-bl}{2\pi a^{3}}.
\label{29}
\end{equation}

So, we can say, that if we define the screw
dislocation by the metric of eq. (20), then, the deflection
angle of light caused by such dislocation is given by eq. (29).

\section{Discussion}

We have defined gravitation as a
deformation of a medium called space-time. We have used
equation which relates Riemann metrical tensor to the
tensor of deformation of the space-time medium and applied
it to the gravitating
system, which we call screw dislocation in space-time. The
term screw dislocation was used as an analogue with
the situation in the continuous mechanics. We derived the angle of
deflection of light passing along the screw dislocation axis at the distance
$a$ from it on the assumption that trajectory length was $l$.
This problem was not considered
for instance in the Will monography [6].The screw dislocation
was still not observed in space-time and it is not clear what role play
the dislocations in the development of universe after big bang.
Our method can be applied
to the other types of dislocations in space-time and there is no problem
to solve the problem in general. We have used here the
specific situation because of its simplicity.
We have seen that the problem of dislocation in space-time is
interesting
and it means there is some scientific value of this problem.
Sooner or later the physics will give answer to the question what is the
the role of dislocations in forming of universe.
\vspace{10mm}

{\bf References}

\vspace{5mm}

[1] I.R. Kenyon, General Relativity (Oxford University
Press Inc. New York, 1996).\\[5mm]

[2] S. Chandrasekhar, Amer. Journal Phys.
40 No. 2(1972) 224.\\[5mm]

[3] A. Tartaglia, Grav. Cosmol. 1 (1995) 335.,
 M. Pardy, gr-qc/0103010.\\[5mm]

[4] L.D. Landau and E.M. Lifshitz, The
theory of elasticity (Mir, Moscow,1987)(in Russian).\\[5mm]

[5]L.D. Landau and E.M. Lifshitz,
The classical theory of fields \\[3mm]\hspace*{5mm} (Pergamon Press, Oxford, 1962).\\[5mm]

[6] C.M. Will, Theory and experiment in gravitational
physics,
Revised edition. \\[3mm]\hspace*{5mm}(Cambridge University Press, Cambridge, 1983).

\end{document}